# Anomalous resistivity upturn in the van der Waals ferromagnet Fe$_5$GeTe$_2$

Yalei Huang, Xinyu Yao, Fangyi Qi, Weihao Shen, Guixin Cao[*]

Materials Genome Institute, Shanghai University, 200444 Shanghai, China

**ABSTRACT**

Fe$_n$GeTe$_2$ ($n$ = 3, 4, 5) have recently attracted increasing attention due to their two-dimensional van der Waals characteristic and high temperature ferromagnetism, which make promises for spintronic devices. The Fe(1) split site is one important structural characteristic of Fe$_5$GeTe$_2$ which makes it very different from other Fe$_n$GeTe$_2$ ($n$ = 3, 4) systems. The local atomic disorder and short-range order can be induced by the split site. In this work, the high-quality van der Waals ferromagnet Fe$_{5-x}$GeTe$_2$ were grown to study the low-temperature transport properties. We found a resistivity upturn below 10 K. The temperature and magnetic field dependence of the resistivity are in good agreement with a combination of the theory of disorder-enhanced three-dimensional electron–electron and single-channel Kondo effect. The Kondo effect exists only at low magnetic field $B < 3\,T$, while electron-electron dominates the appearance for the low-temperature resistivity upturn. We believe that the enhanced three-dimensional electron–electron interaction in this system is induced by the local atomic structural disorder due to the split site of Fe(1). Our results indicate that the split site of Fe plays an important role for the exceptional transport properties.

The quasi-two-dimensional (2D) van der Waals (vdW) bonded materials have captured vast attention due to their twofold merits for both fundamental study and technical applications.[1-5] Many of these materials show promising properties for future applications, such as CrGeTe$_3$,[6] CrI$_3$,[7] and Fe$_3$GeTe$_2$,[8] which have been exfoliated to the monolayer limit. Moreover, the magnetic order in the monolayer limit was found to persist, which makes spintronic devices higher speed and lower energy consumption.[3] The operation of a spintronic device requires ferromagnetism of the materials persists at room-temperature. Thus, such pursuits promote a desire to identify vdW ferromagnetic materials with magnetic order nearby room temperature.

---

[*] Corresponding author, Email: guixincao@shu.edu.cn

As promising vdW ferromagnets, Fe$_n$GeTe$_2$ ($n$ = 3, 4, 5) are very attractive due to the high Curie temperature ($T_C$), where each layer consists of a Fe$_n$Ge slab sandwiched by Te layers.[9-13] Interestingly, Fe$_n$GeTe$_2$ ($n$ = 3, 4, 5) hosts a variety of noticeable effects, namely large anomalous Hall effect,[8,9,14] topological spin texture[15,16] etc. Among the Fe$_n$GeTe$_2$ ($n$ = 3, 4, 5) systems, the Fe$_5$GeTe$_2$ ($R\bar{3}m$) has been widely investigated theoretically and experimentally based on its highest $T_C \sim 270$ K − 363 K.[17-20] Compared with other materials in the Fe$_n$GeTe$_2$ ($n$ = 3, 4) family, the extra Fe layer in Fe$_5$GeTe$_2$ can considerably enhance the magnetic interaction.[10,11] For the structure of Fe$_5$GeTe$_2$, there include three Fe sites per unit cell, Fe(1), Fe(2) and Fe(3) as shown in Fig.1(a). The Fe(1) position is treated as a split site, which it can be at either above or below the neighboring Ge atom, or just be vacant.[11,21] As shown in Fig.1(b) and (c), the possible split-sites of Fe(1) atoms and Ge atoms are labeled brown and green open circles, corresponding to two possible Fe(1)-Ge split sites, Fe(1)$_{up}$ - Ge$_{down}$ and Ge$_{up}$ - Fe(1)$_{down}$.[22] The theory calculations found that a distribution of the split site yields a ferromagnetic (FM) ground state.[23] By using atomic-resolution scanning transmission electron microscopy and single crystal X-ray diffraction, Andrew et al. found that the split-site leads to intrinsic disorder or a short-range order.[9] The scanning tunneling microscopy topography reveals $\sqrt{3} \times \sqrt{3}$ superstructures on the cleavage surface of Fe$_5$GeTe$_2$ due to the short-range order. The obtained $\sqrt{3} \times \sqrt{3}$ ordering of Fe(1)-Ge pair breaks the inversion symmetry, which is an important microscopic origin of the a Dzyaloshinskii-Moriya interaction.[22] Thus, local complicated atomic disorder induced by the split site play a key role in determining the physical properties in Fe$_5$GeTe$_2$. However, so far, the disorder effects on the transport properties of Fe$_5$GeTe$_2$ have been barely investigated.

In this work, we present experimental investigation on the electrical transport and magnetic properties of high-quality vdW Fe$_5$GeTe$_2$ single crystals. We found an emergent resistivity upturn below 10 K. Analytical study indicates that both the electron–electron interaction (EEI) and the single-channel Kondo effect (1CK) cause the correction to the resistivity at low temperatures. The 1CK exists at low magnetic field ($B < 3\,T$) and is completely suppressed when $B \geq 3\,T$. The three-dimensional (3D) EEI dominates the low temperature resistivity upturn due to the enhanced disorder induced by Fe(1) split site.

The high-quality single crystals of $Fe_{5-x}GeTe_2$ were grown by chemical vapor transport (CVT) method with Iodine as the transport. The mixture of Fe (99.99%), Ge (99.99%), and Te (99.99%) with the molar ratio 6:1:2 and iodine (5mg/cm$^{-2}$) were sealed in an evacuated quartz tube, and placed in a tubular furnace. The lowly heated up to 700 °C with a heating rate of 1 °C per minute and maintained for 7 days. Then the furnace was rapidly cooled down to room temperature without any heating. The inset of Fig. 1(d) shows the typical size of the samples. The single crystals were examined by using the X-ray diffractometer (XRD, Bruker D2 PHASER) and Bruker D8 Discovery with home-made high-throughput attachment. The chemical compositions were characterized by energy-dispersive x-ray spectroscopy (EDX, HGST FlexSEM-1000). The crystals utilized in this study have a composition of $Fe_{4.8}GeTe_2$. Magnetization measurements of single crystals were performed in a SQUID magnetometer (MPMS, Quantum Design). The longitudinal resistivity was measured by a standard four-probe method using a Physical Property Measurement System (PPMS, Quantum Design).

Fig. 1(d) shows the XRD pattern collected from an as-grown facet of $Fe_{5-x}GeTe_2$ single crystals. It can be seen that only the (00*l*) Bragg peaks are observed, demonstrating that the exposed surface is *ab* plane. Temperature-dependent magnetization *M(T)* of $Fe_5GeTe_2$ is shown in Fig. 1(e). The magnetization increases sharply as the temperature decreases and crosses the transition temperature at $T_C \sim 290$ K, indicating a paramagnetic to ferromagnetic transition. With further cooling, there are three jumps observed in the *M(T)* curve at $T_1 \sim 275$ K, $T_2 \sim 185$ K and $T_3 \sim 100$ K, which are consistent with previous reports.[11,18] Fig. 1(f) depicts the temperature-dependent resistivity $\rho(T)$ and its first derivative $d\rho/dT$. $\rho$ shows a weak temperature dependent above 100 K. However, the anomalies at $T_C$, $T_1$, $T_2$ and $T_3$ are seen in the $\rho(T)$ curve at the almost same temperature points corresponding to the four transitions in the *M(T)* curves as shown in Fig. 1(f). It is even particularly clear in the $d\rho/dT$ curve. Note that in the previous report, there is only two transitions found *T₁* and *T₃* in the $\rho(T)$ curve.[9,18] Our finding of all four transitions consistent with that appeared in *M(T)* indicates the high quality of the synthesized $Fe_5GeTe_2$ single crystals.

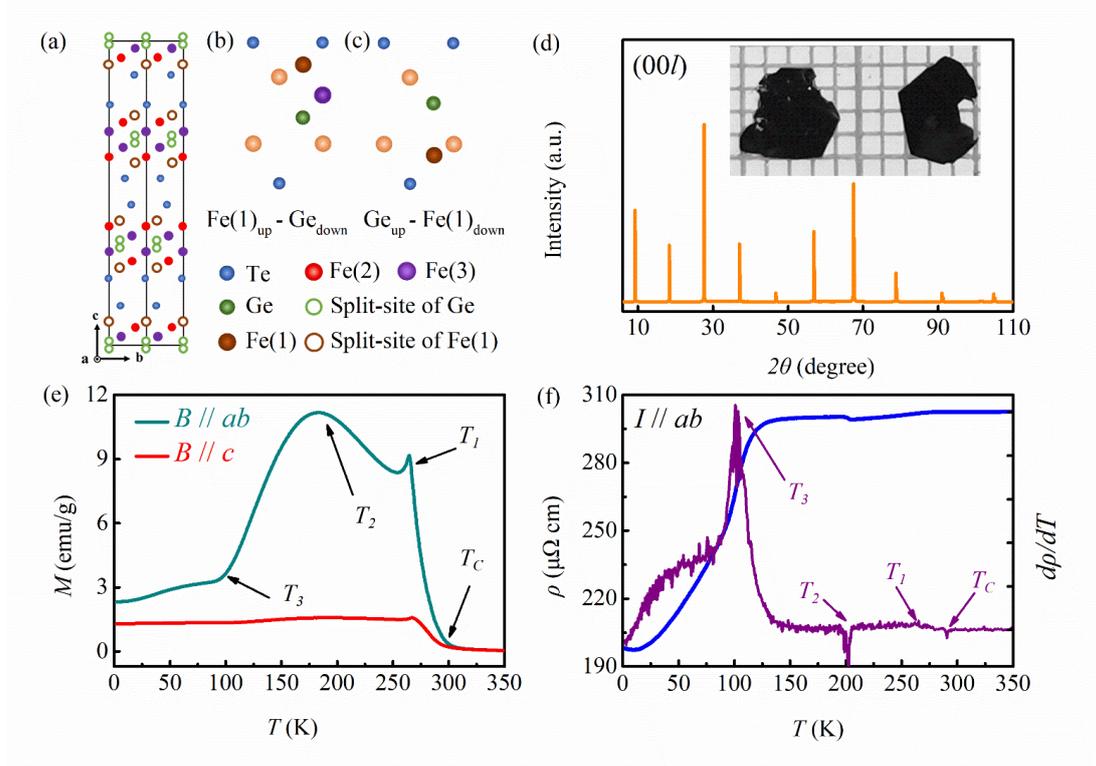

**Fig. 1.** (a) Crystal structure of $Fe_{5-x}GeTe_2$. (b) - (c) Two possible Fe(1) - Ge split-sites of $Fe(1)_{up}$ - $Ge_{down}$ and $Ge_{up}$ - $Fe(1)_{down}$, respectively. (d) X-ray scattering patterns for $Fe_{5-x}GeTe_2$ scanned along the (00$l$) direction. The inset is a photograph of the crystals with the typical size on a millimeter-grid paper. (e) Temperature-dependent magnetization of $Fe_{5-x}GeTe_2$ single crystals under a magnetic field of 100 Oe along the $ab$ plane and $c$ axis. (f) Temperature-dependent resistivity (blue line) and differential resistivity (purple line) of $Fe_{5-x}GeTe_2$ single crystals.

To understand the low temperature (low-$T$) electrical transport properties, it is useful to examine $\rho(T)$ in more detail. For $T \geq 15$ K, the $\rho(T)$ is fitted well by the formula, $\rho = \rho_0 + aT^{5/2}$ as shown in Fig. 2(a). Where $\rho_0$ is the resistivity due to the domain boundaries and other temperature-independent scattering mechanisms. $aT^{5/2}$ term is an empirical fit to the data. The $T^{5/2}$ dependence was reported in other ferromagnets, such as $MnSb_2Te_4$,[24] which is attributed to the contributions from electron-electron, electron-magnon and electron- phonon scattering.[25] Upon further cooling, the resistivity gradually deviates from $T^{5/2}$ dependence. Interestingly, a resistivity upturn appears below $T_{min} \sim 10$ K, where $T_{min}$ is defined as the temperature at which the minimum in the resistivity locates.

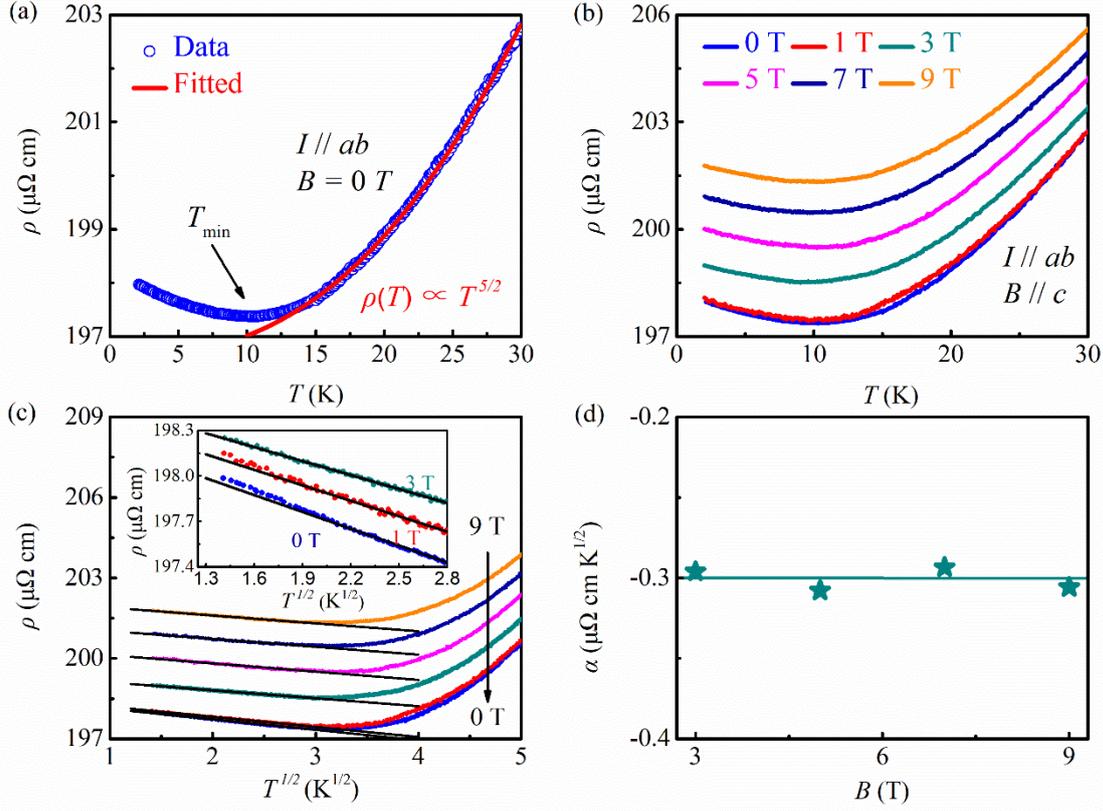

**Fig. 2.** (a) The low-temperature in resistivity of $Fe_{5-x}GeTe_2$ under $I // ab$. The red solid line is the fitted results with $\rho = \rho_0 + aT^{5/2}$ above 15 K. (b) Temperature dependence of resistivity under various magnetic fields (0 - 9 T). (c) The low temperature upturn in resistivity plotted versus $T^{1/2}$ under various magnetic fields. The solid lines are linear fitting of the data. The inset shows the magnified view of temperature-dependent resistivity for 0 T, 1 T and 3T. Except for data at 0 T, all other curves have been vertically shifted for clarity. (d) The dependence of slope $\alpha = d\rho/d\rho(T^{1/2})$ on the magnetic field with the line as guides to the eye.

To investigate in detail low-$T$ resistivity upturn in this compound, we have measured $\rho(T)$ curves under various magnetic fields (0 - 9 T) as shown in Fig. 2(b). It shows that the $T_{min}$ keeps unchanged with the applied magnetic field up to 9 T. Generally, the resistivity upturn at low-$T$ occurs due to the following mechanisms: the EEI,[26] weak localization (WL),[27] and the Kondo effect,[28,29] or the combination of above.[30,31] Based on present data, we discuss in detail the possible origin of the resistivity upturn form EEI, WL and Kondo effect, respectively. We consider firstly the 3D EEI yielding the resistivity upturn. The 3D EEI leads to a $T^{1/2}$ dependence of resistivity because a consequence of the correlation between wave function of the added electron and the wave functions of the occupied electrons that are nearby in energy.[32] We fitted $\rho(T)$ under various fields by $T^{1/2}$ as shown in Fig. 2(c). The fitted results show a good linear dependence on $T^{1/2}$ when $B \geq 3\,T$, with no any sign of saturation as temperature approaches 2 K. While, the magnified view of $\rho(T)$ for 0 T and 1 T displays not well fitting as shown in the inset of Fig. 2 (c). This fitting results indicate the different origin for the low-$T$ resistivity upturn under strong magnetic field ($B \geq 3\,T$) and weak magnetic field ($B < 3\,T$). More strikingly, the slope $\alpha = d\rho/d\rho(T^{1/2})$ is nearly constant as magnetic field varies from 3 to 9 T, $\alpha \sim -3.0\,\mu\Omega\,\text{cm}\,T^{-1/2}$, as shown in Fig. 2(d).

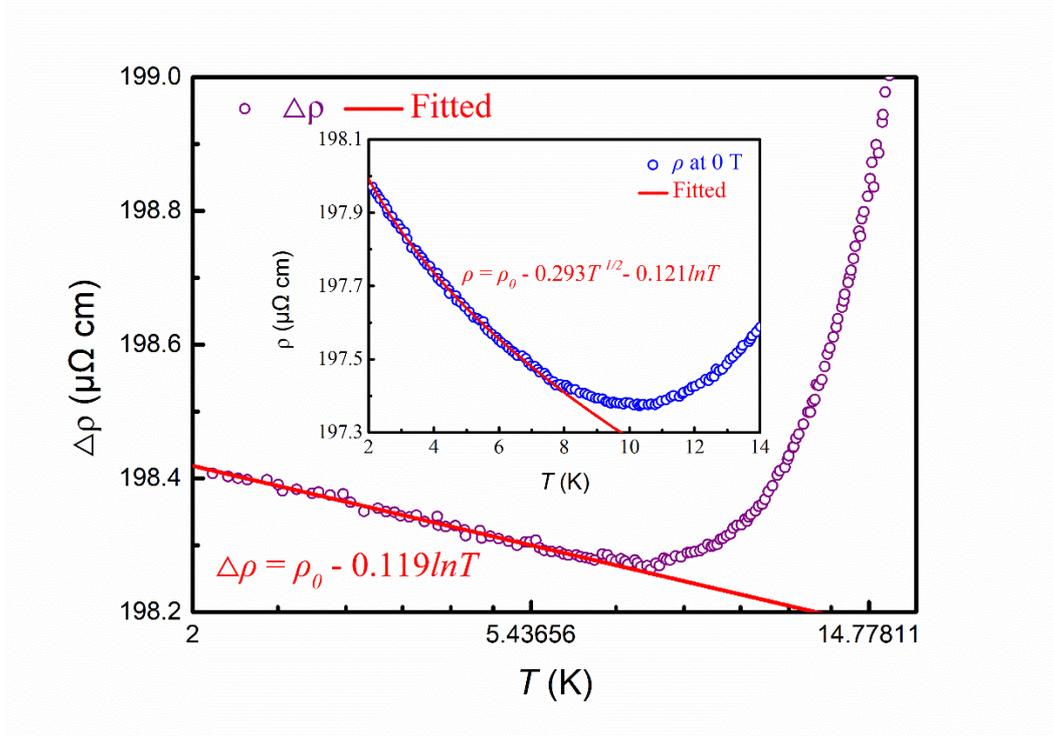

**Fig. 3.** The resistivity of subtracting the contribution of 3D EEI, $\Delta\rho = \rho - \rho_{EEI}$, plotted versus $lnT$ at 0 T, and the solid line is the linear fitting. The inset show that temperature dependence of resistivity at 0 T. The solid line are fitting curves by $\rho = \rho_0 + \alpha T^{1/2} + \beta lnT$.

At first, we discuss the strong magnetic-field case ($B \geq 3\,T$). It should be noted that WL can lead to correction to low-$T$ resistivity, but can be easily suppressed by even weaker magnetic field (on the order of Tesla or less).[33] So, the irrelevance of WL is confirmed by the nearly constant values of $\alpha$ under strong magnetic field. As above, the Kondo effects is also possibly responsible for low-$T$ resistivity upturn. Kondo effect includes spin single-channel Kondo effects (1CK) and two-channel Kondo effects (2CK) in terms of the number of independent channels coupled to a local magnetic moment.[34] We can rule out easily the possible origin form 1CK, which exhibits $\rho \propto lnT$, since our data satisfy $\rho \propto T^{1/2}$. For the 2CK, the orbital 2CK is physically more stable than the spin 2CK, since spin 2CK can scarcely be achieved due to the strict symmetry requirement. The resistivity of orbital 2CK scales with $lnT$, $T^{1/2}$ and $T^2$ in three distinct temperature regimes with decreasing temperature.[35] Our result shows a clear absence of a $lnT$ and $T^2$ dependence, which consequently rules out the possibility from 2CK here. Thus, the 3D EEI is highly consistent with the low-$T$ resistivity upturn of our samples under strong magnetic field ($B \geq 3\,T$).

The above analysis shows that there should exist an additional reason for the low-$T$ resistivity upturn when $B < 3\,T$. Since the 3D EEI is field-independent, the additional contribution for resistivity when $B < 3\,T$ can be obtained by subtracting the contribution of 3D EEI, i.e., $\Delta\rho = \rho - \rho_{EEI} = \rho + 0.3T^{1/2}$. Therefore, we subtract the $\rho(T)$ data under 0 T from 3D EEI and displays the $\Delta\rho$ versus $lnT$ in Fig. 3 which displays a wonderful fitting. The slope $\beta = d\rho/d\rho(lnT) \sim -0.12\,\mu\Omega\,cm\,T$ is obtained. This is consistent with the 1CK. The emergence of a Kondo lattice has been experimentally found in the itinerant ferromagnet $Fe_3GeTe_2$.[36] The Kondo behavior in the $\rho(T)$ at 0 T here is particularly interesting and deserves further detailed investigation. Based on above explanations, we believe that the low-$T$ resistivity upturn under 0 T case should include the contribution from both 3D EEI and Kondo effect. To clear this, we further fit the resistivity upturn under 0 T in the temperature range of 2 - 8 K with the formula, $\rho = \rho_0 + \alpha T^{1/2} + \beta lnT$, as shown in the inset of Fig.3. Here, $\rho_0$ is the residual resistivity, $\alpha T^{1/2}$ term represents the contribution of 3D EEI and $\beta lnT$ arises from the 1CK. From the good fitting resutls, $\alpha \sim -0.29\,\mu\Omega\,cm\,T^{-1/2}$ is obtained, which shows a comparable $\alpha$ value as that fitted for the upturn under 9 T ($\alpha \sim -0.30\,\mu\Omega\,cm\,T^{-1/2}$). Moreover, the obtained $\beta \sim -0.12\,\mu\Omega\,cm\,T$ is also consistent with the fitted $\beta \sim -0.12\,\mu\Omega\,cm\,T$ under $\Delta\rho$. This means that there exists a crossover between 3D EEI and 1CK at low temperature. That is to say, the low-$T$ resistivity upturn was caused by both the 3D EEI and Kondo effect. While the 1CK effect is completely suppressed under an applied field of 3 T. Therefore, the low-$T$ resistivity upturn above 3 T only contributes from 3D EEI. It should be noted that the nearly unchanged $T_{min}$ under various magnetic field as shown in Fig. 2a is because the contribution form 1CK is much less than that from 3D EEI, which cause the negligible change within the measurement accuracy.

Finally, to further understand the origin of the 3D EEI in our compounds, we measured the low-$T$ resistivity of three samples with different degrees of crystal quality. To character the quality of the crystals we used, the residual resistivity ratio (RRR) values was used as a scale. The higher RRR, the less relative degree of the disorder. As shown in Fig. 4(a), the $\rho(T)$ curves for three samples displays that $T_{min}$ increases with decreasing RRR. To obtain the coefficient α of samples with different RRR, we fitted $\rho(T)$ under 9 T by $T^{1/2}$ where the 1CK can be completely suppressed, as shown in Fig. 4(b). The obtained α value as function of $T_{min}$ was shown in the inset of Fig.4(b). The α decreases with increasing $T_{min}$, which reflects the evolution of the 3D EEI strength with structural disorder in different samples. The interaction strength reveals a strong dependence on the structural disorder of the samples, which agrees with the disorder-enhanced 3D EEI.[32] Based on the crystal structure of $Fe_5GeTe_2$, we know that the split site of Fe(1) can induce local atomic disorder.[11,21] We believe that the enhanced 3D EEI in this system is induced by the local atomic structural disorder due to the split site of Fe(1).

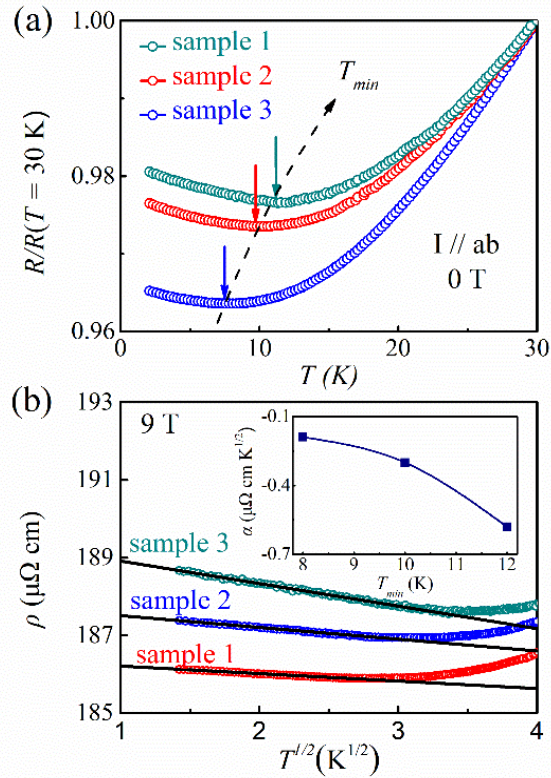

**Fig. 4.** (a) Temperature dependence of resistivity for different samples. (b) The low temperature upturn in resistivity plotted versus $T^{1/2}$ for different samples. Except for data of sample 1, all other curves have been vertically shifted for clarity. The solid lines are linear fitting of the data. The inset shows the dependence of slope $\alpha = d\rho/d\rho(T^{1/2})$ on $T_{min}$ with the line as guides to the eye.

In summary, a resistivity upturn below 10 K was found in van der Waals $Fe_5GeTe_2$ single crystals. The mechanism behind was detailed discussed. Both 3D EEI and 1CK contributes the appearance of low-T resistivity upturn, while 3D EEI play a dominant role. Due to the split site of Fe(1), enhanced three-dimensional electron–electron interaction in this system can be understood. In addition, the finding of Kondo effect in our present compound may point to the Kondo lattice, similar to that reported in $Fe_3GeTe_2$. The correlation between effect of Fe(1) split site and possible Kondo lattice deserve further investigation in the future.

## ACKNOWLEDGMENT

This work was supported by the Ministry of Science and Technology of China (No. 2018YFB0704402). A portion of this work was performed on the Steady High Magnetic Field Facilities, High Magnetic Field Laboratory, CAS.